\documentclass[aps,prl,reprint,groupedaddress]{revtex4-2}

\usepackage{physics}
\usepackage{graphicx}
\usepackage{xcolor}
\usepackage{amsfonts}

\begin{document}

\title{Automated distribution of high-rate, high-fidelity polarization entangled photons using deployed metropolitan fibers}

\author{Alexander N. Craddock}
\author{Anne Lazenby}
\author{Gabriel Bello Portmann}
\author{Rourke Sekelsky}
\author{Mael Flament}
\author{Mehdi Namazi}
\email{Corresponding author. mehdi@quconn.com}
\affiliation{Qunnect Inc., 141 Flushing Ave, Ste 1110, Brooklyn, NY 11205-1005}

\date{\today}

\begin{abstract}
Distributing high-fidelity, high-rate entanglement over telecommunication infrastructure is one of the main paths towards large-scale quantum networks, enabling applications such as quantum encryption and network protection, blind quantum computing, distributed quantum computing, and distributed quantum sensing. 
However, the fragile nature of entangled photons operating in real-world fiber infrastructure has historically limited continuous operation of such networks. 
Here, we present a fully automated system capable of distributing polarization entangled photons over a 34 km deployed fiber in New York City. 
We achieve end-to-end pair rates of nearly $5\times10^5$ pairs/s and entanglement fidelity of approximately $99\%$. 
Separately, we achieve 15 days of continuous distribution, with a network up-time of $99.84\%$.
Our work paves the way for practical deployment of 24/7 entanglement-based networks with rates and fidelity adequate for many current and future use-cases. 
\end{abstract}

\maketitle

\section{Introduction}
Entangled photons not only enhance the current capabilities of quantum links for applications in secure communication \cite{ekert_quantum_1991,Bennett1992-kb}, they are also fundamental for building large scale quantum repeaters \cite{Azuma2023}, distributed quantum computing \cite{monroe_large-scale_2014,Pirandola2016-qm,Wehner2018-vc,Kimble2008-mu}, and distributed quantum sensing networks \cite{eldredge_optimal_2018,Khabiboulline2019-ov,Proctor2018-kr}.
All such use-cases rely on access to high-rate, high-fidelity entanglement across the network. 
Therefore, the high-performance distribution of entanglement over quantum channels is essential for any future quantum network.

Given their weak environmental interactions and speed of propagation, photons are a natural choice as communication qubits.
Among different types of photonic qubits, polarization states provide many advantages such as ease of creation, manipulation, and measurement. 
Such states are also prime candidates for interfacing with atomic and ionic systems \cite{Wilk2007,Stute2012,Craddock2019}, paving the way towards larger scale distributed quantum networks.

Owing to the large amount of unused fiber available and the manageable optical losses over short ranges, telecom fibers are a good choice for photonic quantum channels in metropolitan areas.
However, environmental factors can readily affect fibers and cause changes to the polarization, typically represented as rotations in the Poincaré sphere \cite{Bersin2024}.
These can be both wavelength and time-dependent, and can in turn degrade fiber performance as a quantum channel for entanglement distribution.
As a result, compared to other type of entanglement networks \cite{Pelet2023,Fitzke2022}, the demonstrations of high-fidelity polarization qubit distribution over extended periods of time are difficult \cite{Neumann2022}.
However, due to the advantageous properties of these qubits, much research has been performed on their distribution. 
In the past two decades the field has improved from the early stage experiments over approximately $1$ km fibers \cite{Poppe2004} to inter-country \cite{Neumann2022,Wengerowsky2019} and high rate \cite{Shen2022} links.

\begin{figure*}[t]
    \centering
    \includegraphics[width=\linewidth]{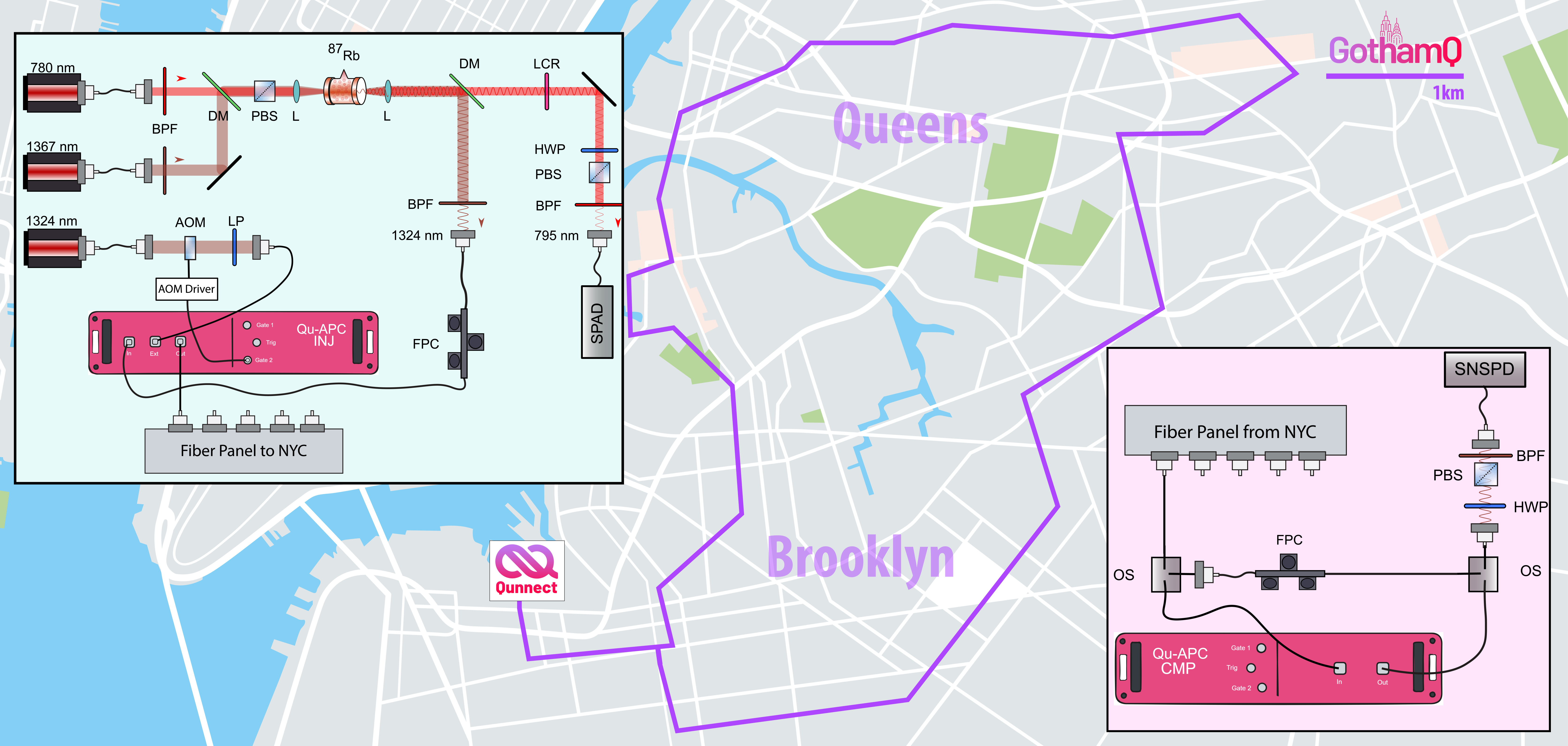}
    \caption{Experimental setup for entanglement distribution experiments.
    Background shows the path of the GothamQ fibers.
    Upper left (lower right) diagram shows the experimental apparatus prior to photons entering (exiting) the GothamQ fiber.
    PBS: polarizing beam splitter, DM: dichroic mirror, BPF: bandpass filter, LCR: liquid crystal retarder, HWP: half waveplate, L: Lens, LP: Linear Polarizer, AOM: acouto-optic modulator, FPC: Fiber polarization compensator, OS: Optical switch, SPAD: Single photon avalanche detector, SNSPD: Superconducting nanowire single photon detector.}
    \label{fig:experimental_setup}
\end{figure*}

Presently, much of the literature has focused on proof-of-principle experiments.
However, practical entanglement networks will require stable, long-term useable polarization entanglement distribution with high-rate, high-fidelity, and high network up-time.
Here, we report a series of measurements performed through buried fiber optics in New York City, using Qunnect's GothamQ testbed shown in Fig. \ref{fig:experimental_setup}.
We characterize the polarization dispersion properties of the fibers in the GothamQ testbed.
Then, using a narrow-band polarization entanglement source and an active polarization-compensating device, we demonstrate entanglement distribution with throughputs of nearly $5\times10^5$ pairs/s and fidelities of approximately $99\%$ over a 34 km long fiber with a total loss of 17.4 dB.
We show that the entanglement distribution system is capable of operating with high performance for over $15$ days without any user input, maintaining a network up-time of $99.84\%$.
To the best of our knowledge, these experiments  are the first showcase of practical and fully automated distribution of entangled photons at high rates and fidelity, enabling a wide range of applications.

\section{Fiber Polarization Dispersion}

\begin{figure}[t]
    \centering
    \includegraphics[width=\linewidth]{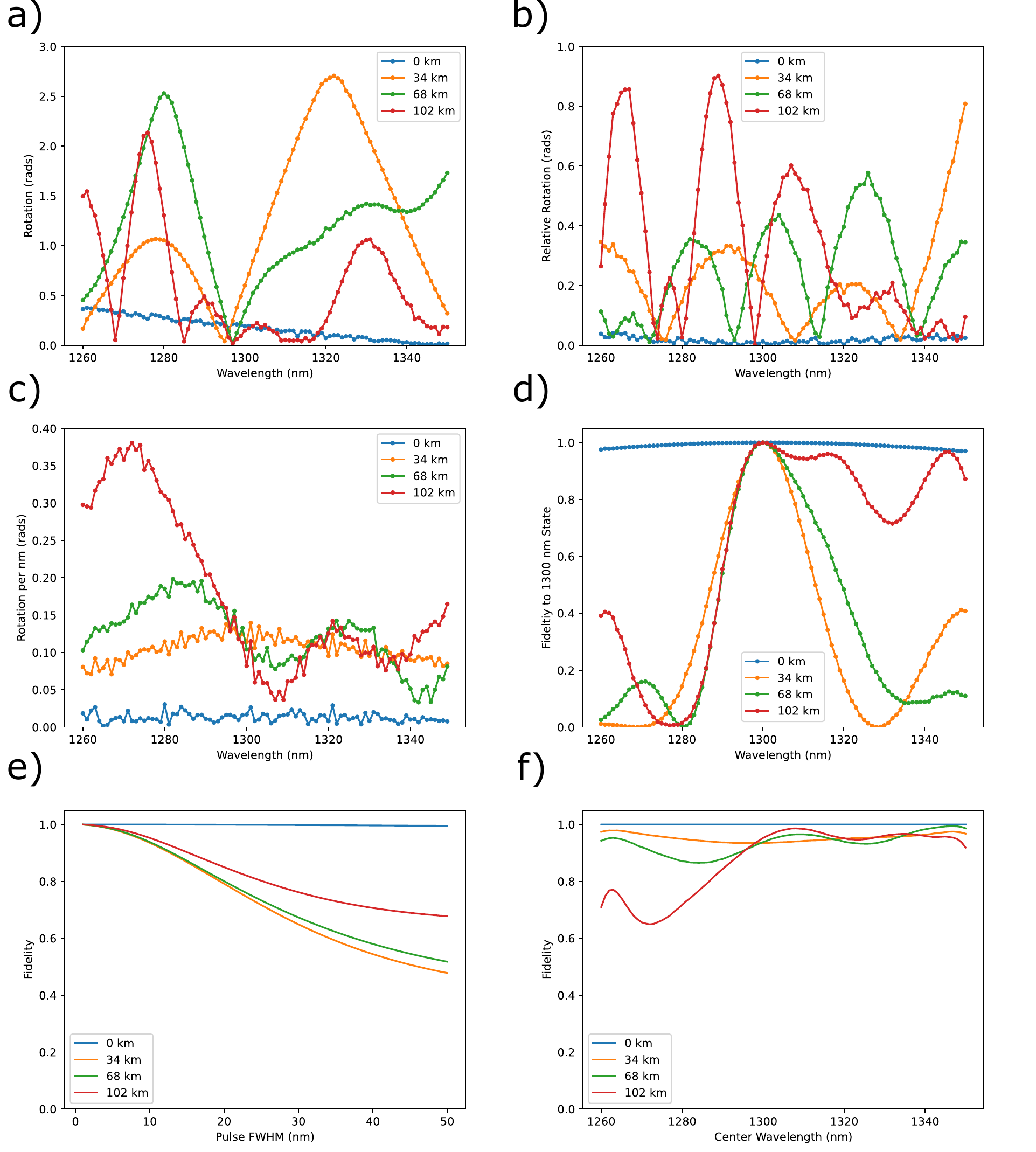}
    \caption{``Instantaneous" fiber polarization dispersion measurements. For each plot the legend indicates the number of GothamQ fibers used for the measurement.
    a) displays the Poincaré sphere rotation as a function of wavelength relative to the input.
    b) shows the Poincare sphere rotation relative to the average rotation value as a function of wavelength.
    c) displays the Poincaré sphere rotation between two wavelengths a nanometer apart.
    d) shows the expected fidelity for a narrowband Bell state where one qubit is passed through the fiber, assuming at the exit of the fiber configuration we have performed an operation to undo the Poincaré sphere rotation for photons at $1300$ nm.
    e) shows the expected fidelity for a Bell state where one photon, with center wavelength of $1300$ nm, is passed through the fiber configuration as a function of the spectral width of the biphoton pair.
    f) displays the expected fidelity for a Bell state for a biphoton pair with a spectral FWHM of $10$ nm, as a function of center wavelength of the photon passed through the fiber.
    In plots a) - d) line is provided to guide the eye.}
    \label{fig:fiber_dispersion_single}
\end{figure}

\begin{figure}[t]
    \centering
    \includegraphics[width=\linewidth]{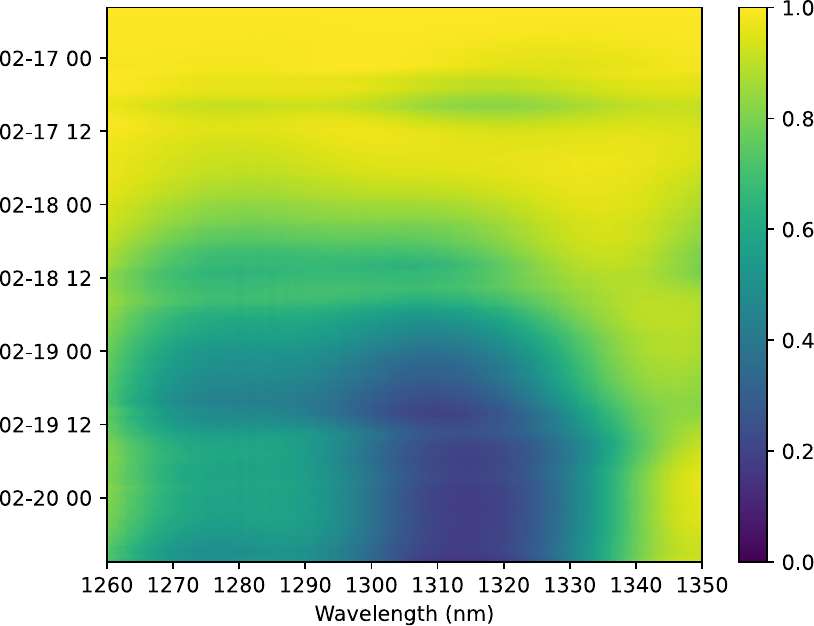}
    \caption{Expected fidelity of a narrowband Bell state over time as a function of the wavelength of the photon passed through one of the GothamQ fibers.
    In each case the fidelity is calculated relative to the initial Poincaré state of the fiber for each wavelength.}
    \label{fig:fiber_dispersion_temporal}
\end{figure}

First, we characterize the polarization dispersion of the fibers in the GothamQ network.
For this work, we use a narrow linewidth laser (Thorlabs TLX3) with a tunable range which covers nearly the entire telecom O-band.
For a given wavelength we generate three polarization states, $\left\{\ket{H},\ket{D},\ket{R}\right\}$.
We pass these through some fiber configuration, and measure the resulting polarization states using a polarimeter (Thorlabs PAX1000IR2).

First we assess the ``instantaneous" polarization dispersion behavior.
For each of the three input polarizations, we sweep the wavelength of the tunable laser from $1260$ nm to $1350$ nm with a $1$ nm step, recording the polarization state at each wavelength.
We perform this measurement for $0$ \footnote{This configuration does not pass through GothamQ, but the light traverses approximately $20$ meters of fiber between common between all the other measurements}, $1$, $2$, and $3$ fibers within GothamQ in series (or distances of $\approx0$, $34$ $68$, and $102$ km respectively).
We note that although the measurement is not truly instantaneous, we have verified that in each case the fiber was stable enough that this approximates an instantaneous measurement.
Knowing the output and input states, we calculate the Poincaré sphere rotation that the fiber configuration performs (see supplementary).
As seen in Fig. \ref{fig:fiber_dispersion_single}a) and b), in all cases we observe some level of dispersive polarization behavior.
To better quantify the dispersion, we calculate the Poincaré sphere rotation per unit nanometer, shown in Fig. \ref{fig:fiber_dispersion_single}c).
From this, we can see clearly that all the configurations which pass through GothamQ fibers exhibit a much higher level of polarization dispersion than the configuration which doesn't.

Given the above fiber behavior, we are able to make some inferences about the suitability of the fibers as quantum channels for distributing quantum entanglement.
In Fig. \ref{fig:fiber_dispersion_single}d), we calculate the fidelity for retaining a Bell state when one qubit is transmitted across a given fiber configuration, assuming we have performed the appropriate correction on the receiving end to account for the Poincaré sphere rotation at $1300$ nm.
By weighting curves such as those in Fig. \ref{fig:fiber_dispersion_single}d) by a Gaussian with a given full-width half-max (FWHM), we are able to make inferences about the fidelity of photon pairs where one photon is propagated along the GothamQ fibers.
In Fig. \ref{fig:fiber_dispersion_single}e) we show the expected fidelity as a function of spectral FWHM of the biphoton pair where the propagating photon has a center wavelength of $1300$, and in Fig. \ref{fig:fiber_dispersion_single}f) we display the expected fidelity as a function of central wavelength of the propagated photon for a biphoton pair with a $10$ nm FWHM.
We see that in all configurations, increasing the pulse bandwidth decreases the expected fidelity of the entangled state.

In addition to the ``instantaneous" measurements, we measure the long-term fiber polarization dispersion behavior, shown in Fig. \ref{fig:fiber_dispersion_temporal}.
For the figure, we calculate the change in fidelity of a hypothetical narrowband Bell state, where one photon is passed through the GothamQ fiber, as a function of wavelength.

From the data it is clear that high fidelity distribution of polarization qubits necessitates that the photons used be narrowband.
Given that some level of polarization drift is observed for every wavelength measured in Fig. \ref{fig:fiber_dispersion_temporal}, maintaining high fidelity polarization qubit transmission in an optical fiber is likely to require active compensation.
However, given our observations in Fig. \ref{fig:fiber_dispersion_single} and \ref{fig:fiber_dispersion_temporal} such compensation needs to be done on the same fiber at the same wavelength as the qubits. 

% Do we wanna have a summing up here?
%comment
\section{Entanglement Distribution}

\subsection{Experimental Setup}

Next, we demonstrate entanglement distribution for pairs of photons where one is distributed in a buried fiber within New York City.
The experimental setup for this is shown in Fig. \ref{fig:experimental_setup}.

We use a spontaneous four-wave mixing (SFWM) source based on warm atomic vapors to generate entangled bichromatic photon pairs.
In the source, a $780$ nm pump and $1367$ nm coupling beam excite atoms within an enriched rubidium-87 vapor cell to the $\ket{6S_{1/2}}$ state via the $\ket{5P_{3/2}}$ intermediate state.
From the doubly-excited state, emission of a $1324$ nm photon may occur, projecting a collective $\ket{5P_{1/2}}$ excitation onto the vapor \cite{davidson_bright_2021}.
Subsequent decay of this collective excitation, which is highly directional, results in a photon at $795$ nm.
Due to the Zeeman structure of rubidium, and the co-linear pump and coupling polarization, the $1324-795$ pairs are emitted in a polarization entangled $\ket{\Phi_+}=\frac{1}{\sqrt{2}}\left(\ket{HH}+\ket{VV}\right)$ Bell state \cite{willis_photon_2011}.
For these experiments we use a source setup similar to the one described in \cite{craddock2023highrate}.
Additionally, we use the optimal values for the pump and coupling detunings, as well as the vapor cell temperature determined in \cite{craddock2023highrate}.
Throughout the following, the pump and coupling powers are set to be equal, and adjusted to achieve the different pair rates shown.
An advantage of this source is that it natively produces narrowband (linewidth less than $1$ GHz) polarization entangled biphotons.
From Fig. \ref{fig:fiber_dispersion_single}, this should mean the fidelity of the entangled state after propagation should not be bounded by the photon linewidth.

To preserve the fidelity of the entangled state as photons are passed through the metropolitan fiber, we use our automated polarization compensation (APC) devices.
The APC operates as a pair, with an injector and a compensator at either ends of the fiber.
The APC injector produces a sequence of classical pulses with well-defined polarization, created using a polarizer followed by an elasto-electro-optic modulator (EEOM).
We pass the classical pulses down the fiber to be stabilized to the APC compensator.
The compensator contains an EEOM (with a modulation bandwidth of approximately $120$ kHz) and a fast polarimeter ($ 10^4$ measurements/s).
We initalize the pair by taking a reference measurement with the polarimeter.
After which, the compensator adjusts its EEOM to ensure subsequent pulse sets from the injector match the reference.
We use time-division multiplexing to mix the classical pulse sequence with light passed to the input port of the injector, via a set of optical switches.
For all the measurements shown, we trigger the APC compensation cycle approximately every $20$ seconds.
In each compensation cycle, the fidelity is first measured by comparing the current state of the fiber to the reference.
If it is found to exceed a trigger threshold, (set here to be $99\%$) then the compensation cycle finishes.
If it is below the threshold, we utilize a gradient descent algorithm to adjust the compensator EEOM until it exceeds an optimization threshold (also set here to be $99\%$).
The compensation cycle takes between $30$ to $1000$ ms, depending on the polarization rotation that has occurred on the fiber.

For this experiment, we couple the $1324$-nm telecom photons produced by the source into a fiber attached to the input port of the APC injector which is in turn connected to the buried $34$-km long metropolitan fiber, shown in Fig \ref{fig:experimental_setup}.
After travelling through the metropolitan fiber, we pass the telecom photons to a fiber switch.
From the switch we either pass the photons to the APC compensator, or a fiber path which bypasses the compensator completely.
A further switch recombines the paths and sends the photons for polarization analysis.
With both switches we are able to toggle between compensating and not compensating for drifts in the metropolitan fiber.

For analysis of the entangled state after fiber transit, we use a measurement station for each of the two source photon wavelengths.
In both cases, the measurement station consists of a half-waveplate and a polarizing beamsplitter.
We perform polarization measurements for the $795$-nm photons prior to fiber coupling at the source setup.
For the $1324$-nm photons, we use an independent measurement station setup that includes a bandpass filter.
This helps to minimize noise that may have been accumulated in the metropolitan fiber.
While full two-photon tomography requires both a quarter- and half-waveplate for both photons for polarization analysis \cite{james_measurement_2001}, we use a method (see supplementary material) that allows us to bound the entangled state fidelity using only measurements of the coincidences in the $\left\{\ket{HH},\ket{HV},\ket{VH},\ket{VV},\ket{DD},\ket{DA},\ket{AD},\ket{AA}\right\}$ modes.

As discussed earlier, close matching of the wavelength of the classical and quantum light is required to ensure good polarization compensation of the fiber for the quantum light.
Therefore, we use a $1324$-nm laser source as our classical light for the APC injector, verified with a wavemeter to be within a nanometer of the signal photons produced by the source.
The laser is left free running throughout the experiment.
After coupling the classical light into fiber, we pass the light through a fiber-based acousto-optic modulator (AOM).
The AOM is on/off controlled by the APC injector to shutter the classical light when active compensation is not occurring.
Along with the in-built optical switches, this minimizes the noise present at the telecom measurement station due to the APC.
After the AOM, we use an in-fiber polarizer to ensure a well-defined input polarization for the classical field passed to the APC injector.

For the $1324$-nm photons, we use a pair of polarization controllers: one prior to the APC injector input and one on the uncompensated path between the two fiber switches.
The initial polarization rotation performed by the metropolitan fiber, as well as other shorter fibers within the setup, is unknown.
We use the fiber polarization controllers, along with a liquid crystal retarder (LCR) on the source setup to undo this initial rotation for the compensated and uncompensated paths.
For the two experiments presented later, the LCR and controllers were calibrated prior to any data being taken, and then the values fixed for the duration of the experiment.

We detect the $795$-nm photons using a set of single-photon avalanche photodiodes and the $1324$-nm photons using a superconducting nanowire single-photon detector (SNSPD) ($\approx 350$-ps and $\approx 90$-ps timing jitters, and $68\%$ and $90\%$ detection efficiencies, respectively).
We find that the metropolitan fiber has a loss of $14.45$ dB.
Along with the other losses due to other elements (see supplementary material), the total loss from the source to the input of the telecom measurement station is $17.46$ dB.

\subsection{Rate vs. Fidelity}

\begin{figure}[t]
    \centering
    \includegraphics[width=\linewidth]{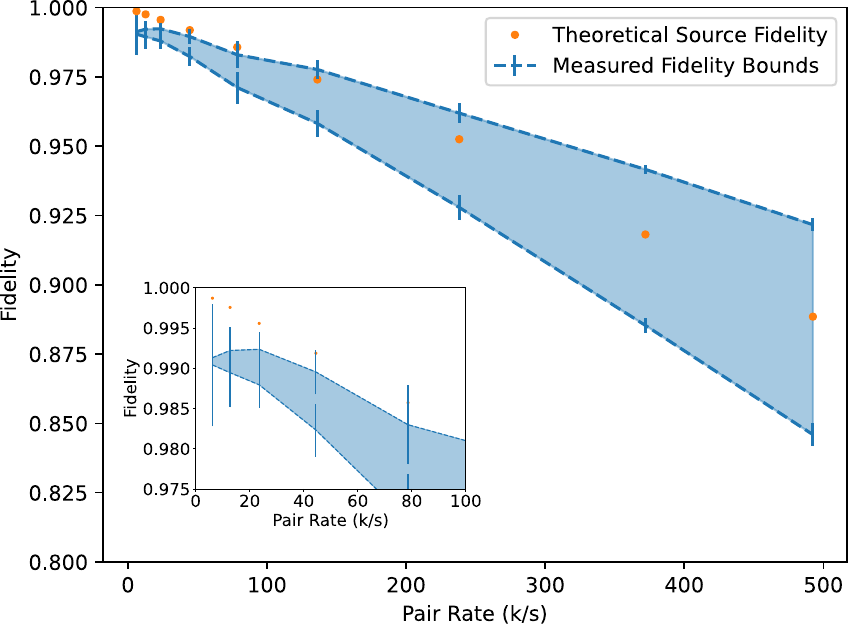}
    \caption{Fidelity to the $\ket{\phi_{+}}=\frac{\ket{HH}+\ket{VV}}{\sqrt{2}}$ Bell state as a function of the throughput pair rate (after $1324$-nm photons are passed through a single GothamQ fiber).
    For each pair rate an hour-long dataset was taken.
    Inset is zoom in of the low rate behavior.
    Fidelity bounds (see supplementary) are the mean over the dataset, error bars are from statistical uncertaities.
    Theoretical fidelity is calculated as $1-\frac{3}{2(1+g_{SI}})$ (see supplementary for derivation).}
    \label{fig:rate_vs_fidelity}
\end{figure}

Here, we explore the dependence of the entanglement fidelity on the fiber pair throughput.
For this experiment, we adjust the pump and coupling powers to change the source entangled pair generation rate. 
Given the probabilistic nature of the entangled pair source used, we expect the pair cross-correlation, and therefore the entangled state fidelity, to decrease as the pair generation rate is increased (see supplementary).
At each generation rate, we take the fidelity-bounding measurements (described above) approximately every four minutes for a total of four hours.
For each fidelity-bounding measurement we calculate the highest lower bound and the lowest upper bound (see supplementary) on the fidelity to the $\ket{\phi_{+}}=\frac{\ket{HH}+\ket{VV}}{\sqrt{2}}$ Bell state.

The average and standard deviation on the lower bounds over the four-hour measurement duration are shown in Fig. \ref{fig:rate_vs_fidelity} for various entangled pair rates.
We note that we can only bound the fidelity due to the measurement method chosen, and that the bounds do not arise purely from experimental uncertainties (see supplementary).
At maximum pump and control power available for this experiment, we can achieve an end-to-end pair distribution rates of approximately $5\times 10^5$.
At this throughput we've experimentally bounded the entangled state fidelity to be $>0.84$.
However, we believe the state fidelity should be source limited to approximately $0.88$.

Reaching such high rates at a high fidelity over real life metropolitan networks opens the path for practical deployment of entanglement-based quantum-secured communication links.
%%Check me please
High fidelity distribution of entangled photons is a key step towards networked quantum computing and distributed quantum sensing over deployed fibers. 
While for high pair generation rates the limit on the fidelity is from the source itself, at lower throughput rates, around $2 \times 10^4$ pairs/s, we can reach a fidelity of approximately $0.99$, which is consistent with the thresholds set for the APC process. 
Achieving such high fidelity for polarization entanglement distribution not only surpasses all previous polarization based deployments, to the best of our knowledge, it is on par with the state-of-the-art for time-bin and time-energy based entanglement networks \cite{Pelet2023,Fitzke2022}.

\subsection{Long-term usability measurement}

\begin{figure}[t]
    \centering
    \includegraphics[width=\linewidth]{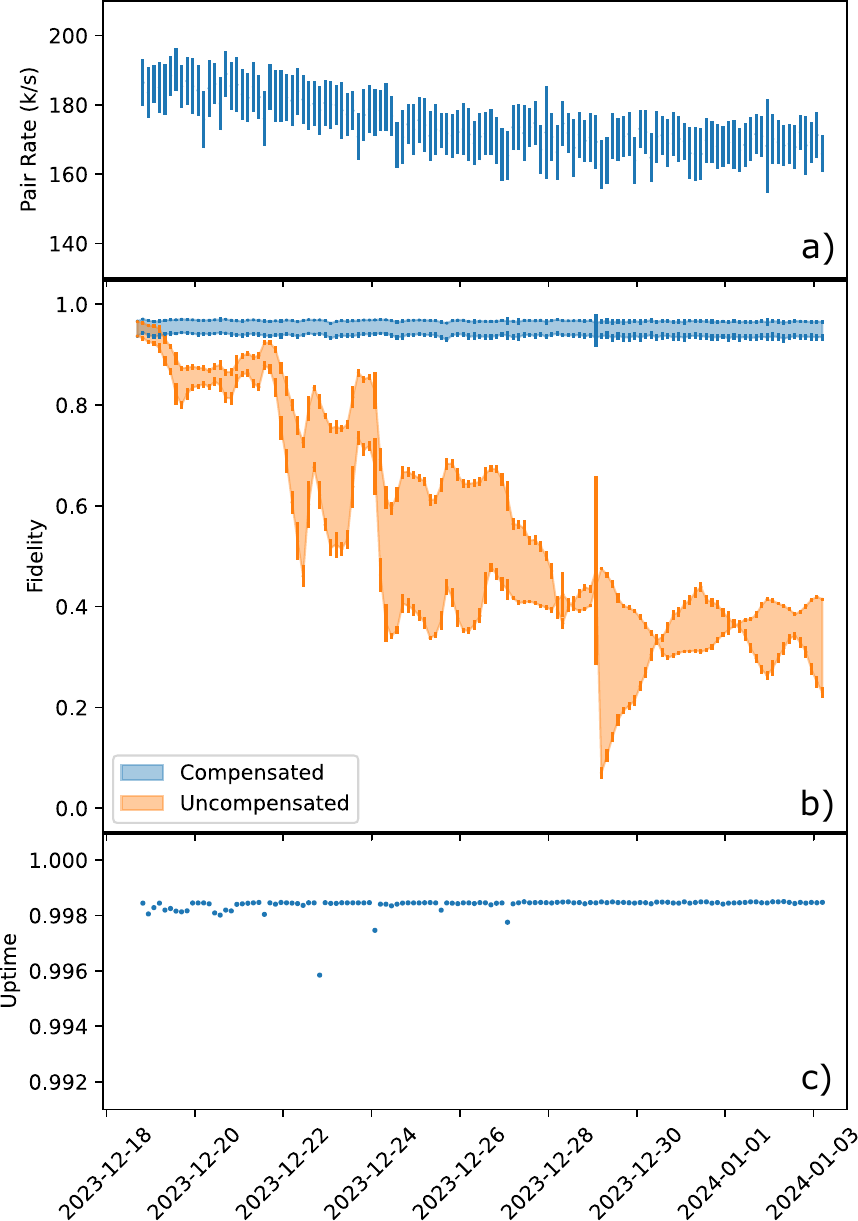}
    \caption{Results of long-term usability measurement.
    a) shows the pair rate observed through the compensated path.
    b) shows fidelity bounds (see supplementary) for the compensated and uncompensated paths.
    c) shows the uptime (time entangled pairs are being actively distributed).
    In both plots we have performed a three-hour trailing average for ease of viewing.
    The uncertainties quoted come from the statistical deviation within the three hour averaging windows.}
    \label{fig:long_boi}
\end{figure}

Another important aspect of a practical quantum network is the long-term, uninterrupted distribution of photon pairs without the need for manual intervention. 
Previous works \cite{Neumann2022} have shown great progress towards this goal but still required relatively time-consuming manual or automatic optimizations which in turn affected the overall network up-time.
Here, we demonstrate the long term usability of the entangled pairs when the $1324$-nm photons from the source are transmitted over the $34$-km metropolitan fiber.
For this experiment the pump and coupling powers is fixed to produce approximately $2\times 10^5$ pairs/s after passing through fiber.
Our experimental loop consists of taking fidelity bounding measurements both for a path compensated by the APC, and an uncompensated path for comparison.
The experimental cycle is repeated approximately every four for over fifteen days.
Similar to the previous experiment, we calculate the highest lower and lowest upper bound on the fidelity of the state, in both the compensated and uncompensated case, to the $\ket{\phi_{+}}=\frac{\ket{HH}+\ket{VV}}{\sqrt{2}}$ Bell state.

In Fig. \ref{fig:long_boi}a) we show the entangled pair rate that passed through the metropolitan fiber.
The pair rate has been calculated by summing the number of coincidences observed in the $\left\{\ket{HH},\ket{VV},\ket{HV},\ket{VH}\right\}$ modes and dividing out peak detection efficiencies.
The observed downward trend in the pair rate over the measurement period was recovered upon completion by adjusting the polarization being sent to the SNSPD.
Therefore, we believe that the actual pair rate was stable over the course of the measurement.

As seen in Fig. \ref{fig:long_boi}b), the state fidelity for the path compensated by the APC is approximately constant over the fifteen-day measurement, with average lower and upper bounds of $0.937 (7)$ and $0.967 (4)$, respectively.
Similar to before, we believe at these rates the fidelity should be limited by the source to approximately $95\%$.
In contrast, the path which is uncompensated experiences significant fluctuations over the course of the measurement, with slow drifts observed in addition to discrete jumps.

In Fig. \ref{fig:long_boi}c) we show the uptime for the data run.
We see that the above results are achieved with a network uptime consistently above $99\%$, with a total uptime of $99.84\%$ over the 15-day run.
To the best of our knowledge, these results are the first ever demonstration of highly stable distribution of high-rate, high-fidelity polarization entangled photons under real-world conditions, opening the path for 24/7 use of quantum entanglement within telecommunication infrastructure.

\section{Conclusion}
The robust distribution of entanglement with high-rate and fidelity across deployed fibers will be critical for the development of the quantum internet. 
Here we presented data from a quantum test-bed in New York City, GothamQ, demonstrating progress towards a fully automated, practical entanglement network. 
Using 34 km of deployed fiber in the city, we showcased the possibility to distribute polarization entangled photons with fidelity up to $99\%$ and rates as high as $5 \times 10^5$ pairs/s. 
Additionally, we presented the long-term usability of the network, achieving high network up-time ($>99.5\%$) for an extended period of operation ($15$ days).
This demonstration shows that robust, high uptime, round-the-clock operation of entanglement distribution networks is attainable for practical use cases.

\section{Acknowledgments}

We thank Thorlabs for generously lending us a TLX3 to perform the fiber polarization dispersion measurements. We thank the Qunnect Inc. team, especially Noel Goddard, Yang Wang, and Felipe Giraldo for their insights and help with manuscript preparation.

\bibliography{main}
\clearpage
\newpage

\onecolumngrid

\renewcommand{\thesubsection}{S.\arabic{subsection}}
\renewcommand{\theequation}{S\arabic{equation}}
\renewcommand{\thefigure}{S\arabic{figure}}
\renewcommand{\thetable}{S\arabic{table}}
\setcounter{equation}{0}
\setcounter{figure}{0}
\section*{Supplementary}

\section{Fidelity Bounding}

Here, we derive a method of bounding the fidelity for a Bell state, given a set of two-photon tomographic measurements in the linear polarization basis.
We consider a general two-photon density matrix
\begin{equation}
    \hat{\rho} = \begin{pmatrix}
\rho_{11} & \rho_{12} & \rho_{13} & \rho_{14} \\ 
\rho_{21} & \rho_{22} & \rho_{23} & \rho_{24} \\ 
\rho_{31} & \rho_{32} & \rho_{33} & \rho_{34} \\ 
\rho_{41} & \rho_{42} & \rho_{43} & \rho_{44}
\end{pmatrix},
\end{equation}
which has a fidelity to the $\ket{\phi_{+}}=\frac{\ket{HH}+\ket{VV}}{\sqrt{2}}$ Bell state
\begin{equation}
    F=\left(\text{Tr}\left[\ket{\phi_{+}}\bra{\phi_{+}}\hat{\rho}\ket{\phi_{+}}\bra{\phi_{+}}\right]\right)^2 = \frac{\rho_{11}+\rho_{44}+\rho_{41}+\rho_{14}}{2}.
\end{equation}
We define a normalization factor
\begin{equation}
    N = C_{\ket{HH}} + C_{\ket{HV}} + C_{\ket{VH}} + C_{\ket{VV}},
\end{equation}
where $C_{\ket{ij}}$ are the coincidences experimentally recorded between the labelled polarization modes.
For the \\$\left\{\ket{HH},\ket{HV},\ket{VH},\ket{VV}\right\}$ basis we can write the on-diagonal elements as
\begin{equation}
\begin{split}
\label{eqn:on_diagonal}
    \rho_{11}&= \frac{C_{\ket{HH}}}{N}\\
    \rho_{44} &= \frac{C_{\ket{VV}}}{N}.
\end{split}
\end{equation}
For the off-diagonal elements, we can make use of the Cauchy-Schwartz inequality
\begin{equation}
    \left|\rho_{ij}\right|^{2} \leq \rho_{ii}\rho_{jj},
\end{equation}
to write
\begin{equation}
\label{eqn:14_inequality}
 -2\sqrt{\rho_{11}\rho_{44}}  \leq \rho_{14} + \rho_{41} \leq 2\sqrt{\rho_{11}\rho_{44}},
\end{equation}
which allows us to bound the fidelity
\begin{equation}
    \frac{C_{\ket{HH}}+C_{\ket{VV}}-\sqrt{C_{\ket{HH}}C_{\ket{VV}}}}{N} \leq F \leq \frac{C_{\ket{HH}}+C_{\ket{VV}}+\sqrt{C_{\ket{HH}}C_{\ket{VV}}}}{N}.
\end{equation}
An additional bound can be found by noting that
\begin{equation}
    \frac{C_{\ket{DD}}+C_{\ket{AA}}}{N} = \frac{1+\rho_{14}+\rho_{23}+\rho_{32}+\rho_{41}}{2}.
\end{equation}
Similar to equation \ref{eqn:14_inequality}, we can write
\begin{equation}
    -2\sqrt{\rho_{22}\rho_{33}}  \leq \rho_{23} + \rho_{32} \leq 2\sqrt{\rho_{22}\rho_{33}},
\end{equation}
and similar to the way we wrote the on-diagonal elements in equation \ref{eqn:on_diagonal} we can write a further set of bounds for the fidelity
\begin{align}
    F \geq &\frac{C_{\ket{HH}} + C_{\ket{VV}} + C_{\ket{DD}} + C_{\ket{AA}} - C_{\ket{DA}} - C_{\ket{AD}} - 2\sqrt{C_{\ket{HV}}C_{\ket{VH}}}}{2N}\\
    F \leq & \frac{C_{\ket{HH}} + C_{\ket{VV}} + C_{\ket{DD}} + C_{\ket{AA}} - C_{\ket{DA}} - C_{\ket{AD}} + 2\sqrt{C_{\ket{HV}}C_{\ket{VH}}}}{2N}.
\end{align}
Due to the rotational symmetry of the Bell state, we can additionally write bounds
\begin{equation}
    \frac{C_{\ket{DD}}+C_{\ket{AA}}-\sqrt{C_{\ket{DD}}C_{\ket{AA}}}}{N} \leq F \leq \frac{C_{\ket{DD}}+C_{\ket{AA}}+\sqrt{C_{\ket{DD}}C_{\ket{AA}}}}{N}
\end{equation}
\begin{align}
    F \geq &\frac{C_{\ket{DD}} + C_{\ket{AA}} + C_{\ket{HH}} + C_{\ket{VV}} - C_{\ket{HV}} - C_{\ket{VH}} - 2\sqrt{C_{\ket{DA}}C_{\ket{AD}}}}{2N}\\
    F \leq & \frac{C_{\ket{DD}} + C_{\ket{AA}} + C_{\ket{HH}} + C_{\ket{VV}} - C_{\ket{HV}} - C_{\ket{VH}} + 2\sqrt{C_{\ket{DA}}C_{\ket{AD}}}}{2N}.
\end{align}

\section{Telecom Photon Loss Budget}

\begin{center}
\begin{tabular}{ |c|c| } 
 \hline
 Element & Loss (dB) \\
 \hline
 Input Paddles & 0.74 \\ 
 APC Injector & 0.22 \\
 Optical Switch & 0.52 \\
 APC Compensator \& Optical Switch & 1.54\\ 
 \hline
\end{tabular}
\end{center}

\section{Expected Entangled State Fidelity From Cross Correlation}

Here, we derive an expression for the expected entangled state fidelity based on the cross correlation function, $g_{SI}$, of a probabilistic pair source.
For the model we assume that the source produces perfectly entangled Bell pairs (without loss of generality we shall assume $\ket{\phi_{+}}=\frac{\ket{HH}+\ket{VV}}{\sqrt{2}}$) in addition to some white noise.
We may write the density matrix for the source output as
\begin{equation}
    \hat{\rho} = a\ket{\phi_{+}}\bra{\phi_{+}}+\frac{1-a}{4}\hat{\mathbb{I}},
\end{equation}
where $a$ parameterizes the noise of the source output and $\hat{\mathbb{I}}$ is the identity matrix.
The entanglement fidelity is calculated as
\begin{equation}
    F= \frac{1+3a}{4}.
\end{equation}
The single polarization mode cross-correlation, $g_{SI}$, can be written as
\begin{equation}
    g_{SI} = \frac{\left<HH\right>}{\left<HV\right>} = \frac{1+a}{1-a}.
\end{equation}
Re-arranging we find the expression quoted in the main text
\begin{equation}
    F=1-\frac{3}{2(1+g_{SI})}
\end{equation}

\end{document}